\documentclass[10pt,pra,amsmath,amssymb,twocolumn]{revtex4}

\usepackage{graphicx}

\begin{document}

\title{Potential energy surface of the ${1}^2A'$ Li${}_2+$Li doublet ground state}

\author{Jason N. Byrd}
 \email{byrd@phys.uconn.edu}
 \affiliation{Physics Department, University of Connecticut, Storrs, CT 06269}
\author{John A. Montgomery, Jr.}
 \affiliation{Physics Department, University of Connecticut, Storrs, CT 06269}
\author{H. Harvey Michels}
 \affiliation{Physics Department, University of Connecticut, Storrs, CT 06269}
\author{Robin C\^{o}t\'{e}}
 \affiliation{Physics Department, University of Connecticut, Storrs, CT 06269}

\begin{abstract}
The lowest doublet electronic state for the lithium trimer ($1^2A'$) is
calculated for use in three-body scattering calculations using the valence
electron FCI method with atomic cores represented
using an effective core potential.  It is shown that an accurate description of
core-valence correlation is necessary for accurate calculations of molecular bond lengths, frequencies and
dissociation energies.
Interpolation between $1^2A'$ {\em ab initio} surface data points in a sparse
grid is done using the global interpolant moving least squares method with a
smooth radial data cutoff function included in the fitting weights and bivariate polynomials
as a basis set.  
The Jahn-Teller splitting of the $1^2E'$ surface into the $1^2A_1$ and $1^2B_2$
states is investigated using a combination of both CASSCF and FCI levels of
theory.  Additionally, preliminary calculations of the $1^2A''$ surface are also
presented using
second order spin restricted open-shell M{\o}ller-Plesset perturbation theory.
\end{abstract}

\maketitle

\section*{Introduction}

With the success of ultracold molecular formation among the alkali metals over the
past decade via photoassociation \cite{jonesphoto} and more recently with Feshbach resonances
\cite{kohlerfesch}, the dynamics of molecules in ultracold traps have
become an important topic to many physicists.  Alkali dimers have been formed in many
different homonuclear and heteronuclear configurations, for both singlet and triplet
ground electronic states.  Furthermore, recent experiments using KRb
\cite{djin1,djin2} and theoretical proposals for LiNa \cite{philippe1} have shown the
possibility of efficiently forming ultracold ground vibrational state diatoms.  
While both ground and excited singlet and triplet states of alkali diatoms have been
studied extensively both experimentally and theoretically, alkali trimers have generally
been ignored by both theorists and experimentalists alike.  In the last few years there has
been an increase of interest in few-body physics with continued success in
the cooling and trapping of atoms.  

In these few-body systems, experimental interest in three body effects range from loss
rate predictions to probing few-body quantum effects such as Efimov states
(Esry and co-workers \cite{greene} and Grimm and co-workers \cite{rgrimm1} for
example) due to the strong non-additive effects
seen in alkali systems.  To date, {\em ab initio} calculations for the sodium trimer
have been done by several groups \cite{na1,na3,na4}
as well as the potassium trimer \cite{na1,k1,hauser}.  In the case of lithium, the quartet 
ground state $1^4A'$ surface has been well studied
\cite{pack1,parker1,cvitas1,parker2} whereas the doublet system has been effectively ignored.
The Jahn-Teller effect and conical intersections between the $1^2E'$, $2^2E'$ and $1^2A'$
surfaces have been studied \cite{gerber,sadygov}, as well as the vibrational \cite{meyer1},
rovibrational \cite{meyer2} and hyperfine \cite{meyer3} structures of the lithium
trimer.  However, to the best of our knowledge the entire ground or excited state
surfaces for the doublet lithium trimer have not been completely studied.

The structure of this paper is as follows, we first discuss the inclusion of
core polarization potentials to accurately describe the effects of core valence
correlation while using an effective core potential (ECP).  The steps taken to
optimize the basis set so as to provide accurate benchmark Li${}_2$ spectroscopic
values are then described in detail.  Following this we show how accurate calculations
of the potential energy surface (PES) were done with a low density of {\em ab initio} points
through the use of the global interpolant moving least squares (IMLS)
fitting procedure.  Finally we describe the calculations we have done
on both the $1^2A'$ and $1^2A''$ surfaces of the lithium trimer and future goals
for the use of these surfaces.  In this work, all calculations have been done using
the MOLPRO 2008.1 quantum chemistry package \cite{molpro} unless otherwise stated.

\section*{Computational Details}

\subsection*{Core-Valence Correlation}\label{cvc}

To accurately describe dissociation energies, equilibrium geometries and vibrational
frequencies in alkali-metal clusters, it is necessary to account for the
electronic core-valence (CV)
correlation energy \cite{JMartin1}.  For all electron calculations this is possible
for the lighter alkali atoms (Li through K) by using the explicitly
correlated basis of Iron, Oren and Martin (IOM) \cite{ironmartin}.  This
approach has recently been done by Cvita\v{s} {\em
et al} \cite{cvitas1} for the spin aligned ${1}^4A'$ Li${}_3$ surface
using spin restricted coupled cluster calculations with single, double and iterative triple
excitations (RCCSD(T)).  To account for CV correlation in heavier atoms, where all
electron calculations are prohibitively expensive, it is necessary to use a
core polarization potential (CPP).  This is also necessary for 
valence electron full configuration interaction (FCI) calculations where an ECP
(a physically equivalent representation to the frozen core
approximation which has no CV correlation) is substituted for the atomic
electrons.  This method has been used with great success for
calculating both lithium dimer potential curves \cite{jasik} and trimer potential
surfaces of potassium \cite{hauser}.

The theoretical description of the CPP is a straightforward addition to that of the ECP
model of atomic cores.
In the Born-Oppenheimer approximation the non-relativistic molecular Hamiltonian can
be separated into kinetic and interaction operators $T$ and $V$ respectively.
Approximating the core of each nuclei with an $l$ dependent pseudopotential and
including the polarization effects at the nuclei gives for the interaction operator
\begin{equation}
V = \sum_k (V^{k} + V^{k}_{cpp}) + \sum_{j>i} \frac{1}{r_{ij}} + V_{cc},
\end{equation}
where
\begin{equation}
V^k = \sum_i -\frac{Q_k}{r_{ik}} + \sum_{il} B^k_{il} exp(-\beta^k_{il}r^2_{il})
P^k_l
\end{equation}
is the core pseudopotential,
\begin{equation}
P^k_l = \sum_m |k l m_l \rangle \langle k l m_l|
\end{equation}
is the projection operator onto the subspace of angular momentum $l$ on core
$k$ and $V_{cc}$ is the core-core coulomb interaction.  Here the
polarization potential, $V^k_{cpp}$ for a given core $k$, is expressed in terms of a static
polarizability and external field at the nuclei position by 
\begin{equation}
V^{k}_{cpp} = -\frac{1}{2}\alpha_k {\bf F}_k\cdot{\bf F}_k.
\end{equation}
where the electric field ${\bf F}_k$ at core $k$ arising from the coulomb
interaction with the electrons
at ${\bf r}_{ki}$ and other cores at ${\bf R}_{kj}$ is 
\begin{equation}
{\bf F}_{k} = \sum_i \frac{{\bf r}_{ki}}{r^3_{ki}}C(r_{ki}) - \sum_j \frac{Z{\bf
R}_{ki}}{R^3_{ki}}.
\end{equation}
The value of the static polarizability for Li${}^+$ is
$\alpha_k=0.1915a_0$ \cite{stuttgartcpp} and the cutoff function $C(r_{ki})$ defined
by Fuentealba {\em et al} \cite{stuttgartcpp} is given as
\begin{equation}
C(r_{ki}) = (1-e^{-(\delta_k r_{ki})}),
\end{equation}
with the cutoff parameter chosen to be $\delta_k=0.831a^{-2}_0$ \cite{stuttgartcpp}.
This form of the cutoff parameter was first presented by M\"{u}ller {\em et al}
\cite{muller} and produces good agreement for ground and low excited states, however
it does show diminished results for Rydberg states \cite{muller}.  Our
calculations using this core polarization potential to describe the
core valence correlation energy were done using the
MOLPRO 2008.1 \cite{molpro} implementation of the Fuentealba {\em et al} \cite{stuttgartcpp} CPP.  

\subsection*{Basis Set}\label{bs}

\begin{table}[!h]
\caption{\label{basisset}
Uncontracted basis set exponents for for the lithium Stevens, Basch and
Krauss \cite{sbk} pseudopotential basis used in this work.  The $s$, $p$ and $d$
orbital exponents are each scaled to give an optimal dissociation energy for
Li${}_2$ as discussed in the text.}
\begin{ruledtabular}
\begin{tabular}{c|ccc}
Orbital Type & SBK       & LFK       & Scaled LFK\\
$s$ & 0.6177000 & 0.6177000 & 0.52504500\\
  & 0.1434000 & 0.1434000 & 0.12189000\\
  & 0.0504800 & 0.0504800 & 0.04290800\\
  & 0.0192300 & 0.0192300 & 0.01634550\\
$p$& 0.6177000 & 0.6177000 & 0.00690155\\
  & 0.1434000 & 0.1434000 & 0.64858500\\
  & 0.0504800 & 0.0504800 & 0.15057000\\
  & 0.0192300 & 0.0192300 & 0.05300400\\
  &           & 0.0065729 & 0.02019150\\
$d$&           & 0.1346266 & 0.13799227\\
  &           & 0.0546860 & 0.05605315\\
  &           & 0.0180355 & 0.01848639\\
  &           & 0.0076882 & 0.00788041
\end{tabular}
\end{ruledtabular}
\end{table}

\begin{table*}[hbt]\begin{ruledtabular}\begin{tabular}{lcccc}
Method/Basis & 
$r_e~$[\AA]   & $\omega_e~$[cm${}^{-1}$] & $D_0~$[cm${}^{-1}$] & $D_e~$[cm${}^{-1}$] \\
 \hline
Expt.\footnote{Taken from Herzberg \cite{herzberg}} & 
2.673 & 351.43 & 8434.58 & 8516.36\footnote{Extracted from the scattering RKR data
\cite{robinrkr}}\\
Recommended\footnote{
FCI/SBK+CPP with scaled LFK using $\beta_s=0.85,$ $\beta_p=1.05$ and $\beta_d=1.025$}
& 2.667 & 352.98 & 8340.12 & 8516.43\\
   FCI/SBK  scaled LFK & 2.693 & 346.54 & 8198.85 & 8371.83\\
       FCI/SBK+CPP LFK & 2.663 & 353.89 & 8338.45 & 8515.10\\ 
           FCI/SBK LFK & 2.687 & 347.19 & 8197.18 & 8371.04\\
RCCSD(T)/CVQZ\footnote{ CVnZ is the n-zeta core valence correlation consistent basis 
set from IOM \cite{ironmartin}.} & 
2.676 & 353.05 & 8294.15 & 8470.27\\
RCCSD(T)/CV5Z\footnotemark[1] & 2.674 & 353.08 & 8311.70 & 8488.46
\end{tabular}\end{ruledtabular}
\caption{\label{spectrocomp}Comparison of different spectroscopic constants of
Li${}_2$ using benchmark basis sets with both explicit inclusion of core-valence
correlation in RCCSD(T) calculations or through the empirical contribution
through a core polarization potential with valence electron full configuration
interaction.}\end{table*}

Atomic and molecular polarizabilities are important factors for long range
interatomic and molecular interactions.  In ultracold systems this is a dominant contribution to
the scattering length in addition to the location of the inner wall.  To accurately
describe both of these quantities in the lithium trimer, we require that the basis
set be flexible enough to describe atomic polarization at long range while
accurately representing the inner wall.
Very accurate polarizability calculations require an accurate description of
electron correlation
and a large basis set containing diffuse functions such as
the Sadlej basis sets \cite{sadlej1,sadlej2}, but this is computationally
impractical to implement for large all electron systems.  Accurate polarizability
calculations can be achieved using an ECP with a small but well chosen basis set as
demonstrated by Labello, Ferreira and Kurtz (LFK) \cite{LFK1,LFK2}.  Their basis
set augments the Stevens, Basch and Krauss (SBK) ECP basis \cite{sbk}, with an additional
$p$ function and four extra $d$ functions optimized following the Sadlej
\cite{sadlej1,sadlej2} method.  

\begin{figure*}[!ht]
\resizebox{5in}{!}{\includegraphics{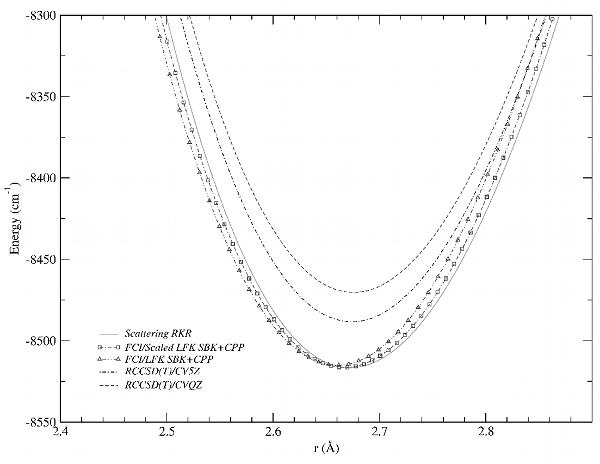}}
\caption{\label{basischeck}
Benchmark singlet Li${}_2$ potential energy curves using both the core-valence correlation
consistent basis sets of Iron, Oren and Martin \cite{ironmartin} and core
polarization potential Stevens, Basch and Krauss pseudopotential \cite{sbk}
with an extended and scaled ECP  basis set calculated at the RCCSD(T) and full
configuration interaction level of theory respectively.  The scattering RKR
curve is the inner wall shifted version of C\^{o}t\'{e} {\em et al} \cite{robinrkr}
such that the potential predicts the correct scattering lengths and Feschbach
resonances.}
\end{figure*}

Using the SBK pseudopotential, we further optimized the uncontracted LFK basis (see Table
\ref{basisset} for the exponents) with and without the CPP for the
Li${}_2$ ${X}^1\Sigma^+_g$ ground state.
This was done using three scale factors
$\beta_{\lambda}(\lambda=s,p,d)$
optimized at the FCI level to give the best calculated value of the 
dissociation energy.
As a benchmark, the CVQZ and CV5Z IOM basis \cite{ironmartin} at
the CCSD(T) level correlating all electrons (no frozen core) gives an error in
the calculated dissociation energy of $46.087$ cm${}^{-1}$ and $27.902$ cm${}^{-1}$
respectively, while at the valence FCI level the unscaled LFK basis
the error is $1.25$ cm${}^{-1}.$  Optimizing the $\beta_\lambda$ coefficients to give the best
dissociation energy, we obtain a difference of $0.077$ cm${}^{-1}$ using the
scaled exponents listed in Table \ref{basisset}, corresponding to the scale
factors
$\beta_s=0.85,$ $\beta_p=1.05$ and $\beta_d=1.025.$  In Figure \ref{basischeck} and
Table \ref{spectrocomp} the
results from the different basis functions and
methods can be seen compared to the Rydberg-Klein-Rees (RKR) curve adjusted to
reproduce the ultracold scattering results \cite{marko}.
We chose to use this scaled, uncontracted LFK basis with the SBK ECP and
Fuentealba CPP in calculating the potential
energy surface of the lithium trimer at the FCI level of theory.

\subsection*{Surface Representation}\label{imls}

Calculations of {\em ab initio} points at the FCI level are very computationally
intensive, even for three electrons with the compact basis just described, so to accurately
describe an entire potential energy surface with a low density of {\em
ab initio} points we implemented the global IMLS fitting
method \cite{lancaster}.  In this method the potential energy at an arbitrary point
$Z$ in the $(x,y)$ plane is approximated by the use of a linearly independent basis
$b_j(Z)(j=1,\dots,n)$ and expansion coefficients $a_j(Z)(j=1,\dots,n)$ so that the interpolated energy is
given by
\begin{equation}
V_{int}(Z) = \sum^{n}_{j} a_{j}(Z)b_{j}(Z).
\end{equation}
The expansion coefficients ${\bf a}(Z)$ are found by minimizing the error function
\begin{equation}
E(V_{int}) = \sum^{N_d}_{j=1}w_j(Z) \left(\sum^{n}_{i=1} a_i(Z)b_i(Z) -
f_j(\ell_j)\right)^2
\end{equation}
of the interpolated energy $V_{int}$ and the {\em ab initio} energy
$f_i(\ell_i)(i=1,\dots,N_d)$ at coordinates $\ell_i$ where $N_d$ is the number of {\em ab initio} data points.  
\begin{figure}[!ht]
\resizebox{3in}{!}{\includegraphics{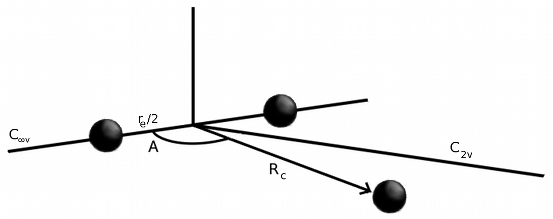}}
\caption{
\label{geometry}
Molecule frame Jacobi coordinates are used to describe the lithium trimer
geometry.  Within this coordinate system the potential energy surface is calculated
assuming that the lithium dimer bond $r_e$ is adiabatically relaxed as the colliding
lithium atom approaches.}
\end{figure}

Expressing the solution to the normal equations $\partial E(V_{int})/\partial a_j=0$
in matrix form we obtain the linear equation for the expansion
coefficients \cite{lancaster}
\begin{equation}\label{equationbla}
B W(Z)B^T{\bf a(Z)} = B W(Z){\bf f},
\end{equation}
where $W(Z)$ is the diagonal matrix of weights $w_i(Z)$ and $B$ is the matrix
\begin{equation}
B=\left(
\begin{matrix}
b_1(\ell_1)& b_1(\ell_2) &\cdots &b_1(\ell_{N_d}) \\
b_2(\ell_1)& \ddots   &       &b_2(\ell_{N_d}) \\
\vdots  &          &\ddots &\vdots       \\
b_n(\ell_1)& b_n(\ell_2) &\cdots &b_n(\ell_{N_d})
\end{matrix}\right).
\end{equation}
The linear system in Eq. \ref{equationbla} is routinely ill conditioned
and so is solved by singular value decomposition.

The weights $w_i(Z)$ dictate the effective range at which a given {\em ab initio}
point will contribute to the global fit and the effective contribution to the
fit.  We use Guo {\em et al}'s \cite{guo1} form of the weight
function, which introduces a cutoff function $S(\chi)$ to the unnormalized weight
function $v_i(\|Z-\ell_i\|)$ so as to smoothly go to zero at a given cutoff radius
$R_{cut}.$  The cutoff function is given \cite{guo1} by
\begin{equation}
S(\chi) = 
\begin{cases}
(1-\chi^m)^4 & \text{if}~0\le \chi\le1,\\
0 & \text{if}~\chi> 1,
\end{cases}
\end{equation}
with $m=10$ and the unnormalized weight function is
\begin{equation}\label{otherbla}
v_i(Z) = \frac{exp[-\|Z-\ell_i\|^2/\zeta^2]}{(\|Z-\ell_i\|/\zeta)^4 +\epsilon}
\end{equation}
where $\epsilon=10^{-10}$ removes the singularity at $\ell_i$. 
Then the normalized weight function is 
\begin{equation}
w_i(Z) = \frac{S(\|Z-\ell_i\|^2/R_{cut})v_i(Z)}{\sum^{N_d}_j
S(\|Z-\ell_i\|^2/R_{cut})v_j(Z)},
\end{equation}
where $R_{cut}$ is to be determined as to give the best fit.
Finally the basis functions are chosen to be bivariate polynomials of order
$n=6$ such that 
\begin{equation}
b(Z) =1,Z_1,Z_2,Z_1^2,Z_2^2,Z_1Z_2,\dots,Z^{n-1}_1Z_2,Z_1Z^{n-1}_2,
\end{equation} 
where the inverse coordinates 
$Z_i = 1/x_i$ are used in this work.  
With the choice of coordinates used here
there is a coordinate singularity in $C_{2v}$ symmetry.  To avoid
this all coordinates $x_i$ are shifted by the same positive, arbitrary additive factor for
the fit then transformed back upon completion.  
The scaling parameter $\zeta$ was chosen to be the average distance between the
interpolant point $Z$ the {\em ab initio} points.  With this definition the cutoff radius was
defined in terms of $\zeta$ as $R_{cut} = 50.0*\zeta$.
This interpolant method is used for the
lithium trimer $1^2A'$ PES to obtain a
global fit using a low number of {\em ab initio} points as references.

\section*{Results and Discussion}\label{ending}

\begin{figure*}[!ht]
\resizebox{4in}{!}{\includegraphics{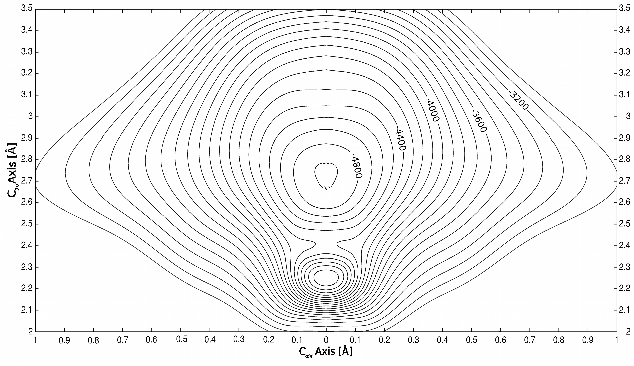}}
\caption{
\label{ap}
Near equilibrium geometry potential energy surface for the Li${}_3~$ $1^2A'$
electronic state calculated using valence elctron full configuration interaction
theory.  Equilibrium
is found to be at bond lengths of $r_e=3.218$\AA~and $R_c=2.238$\AA~for $C_{2v}$
geometry configuration.}
\end{figure*}

The $1^2A'$ surface of Li${}_3$ was calculated at the full configuration
interaction level using the scaled LFK basis set, the SBK pseudopotential
\cite{sbk} and core polarization potential described above, with the three valence
electrons included in the FCI calculation.  
At the FCI level, there are $410670$ configurations of  $A'$ symmetry and
$383292$ configurations of $A''$ symmetry.
All FCI calculations were performed
with the Knowles-Handy determinant FCI program \cite{fci1,fci2} using the MOLPRO 2008.1
package \cite{molpro}.
The geometry was chosen so to best describe the diatomic-atomic collision process.  As
such we used the molecular frame Jacobi coordinates for a homonuclear system
where we define a vector $r_e$ along the diatomic inter-nuclear axis and
another vector $R_c$ from the diatomic center of mass to the colliding atom where the
collision angle $A$ is defined from the $C_{\infty v}$ axis (see Figure
\ref{geometry}).
With this coordinate system the most efficient grid of {\em ab initio}
points is an evenly spaced angular grid with the radial grid chosen to have
the highest density of points at the minima.  We calculated $26$
with the collision angle varying between $90^\circ$ and $45^\circ$ on the
$1^2A'$ Li${}_3$ surface by choosing $A$ and $R_c$ then optimizing the dimer bond
length to give the lowest energy configuration.
The $1^2A'$ state is found to have a triangular equilibrium geometry on
the $1^2A_1$ surface with $r_e=3.218$\AA~ and $R_c=2.238$\AA~for the Jacobi bond
lengths with a dissotiation energy of $4979.42$~cm${}^{-1}$.  The interpolated
surface near the equilibrium geometry is show in Figure
\ref{ap}.

\begin{figure}[!h]
\resizebox{2.5in}{!}{\includegraphics{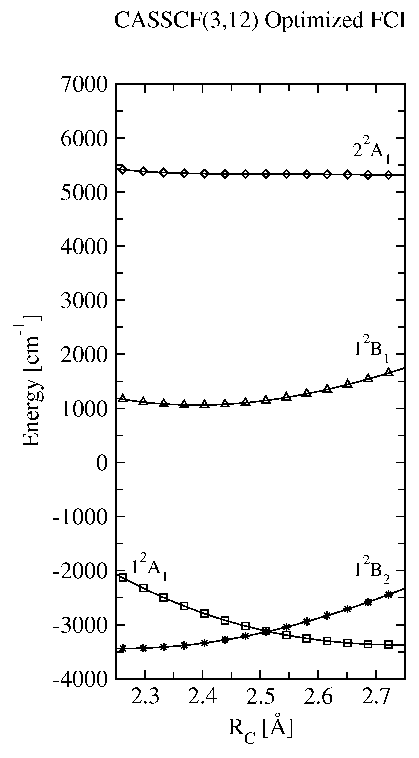}}
\caption{
\label{conical}
First four Li${}_3~$ doublet states in $C_{2v}$ symmetry near the $1^2A_1$ state
equilibrium.  A conical interesction is observed at $R_c=2.51$\AA~between
the $1^2A_1$ and $1^2B_2$ states with a further Li${}_3~$ minimum seen near
$R_C=2.25$\AA~for the $1^2B_2$ state.}
\end{figure}
We have investigated the lowest states of ${}^2A_1$, ${}^2B_1$ and ${}^2B_2$
symmetry and the first excited ${}^2A_1$ state in point group $C_{2v}$
using the complete active shell (CAS) method in conjunction with
FCI.  The same SBK and CPP representation of the core as in the FCI
calculation of the $1^2A'$ surface was used here with an active space of $12$
orbitals ($5a_1 2b_2 4b_2 1a_2$) for the energies of
the $1^2A_1$, $1^2B_1$, $1^2B_2$ and $2^2A_1$ states as seen in Figure
\ref{conical}.  Here $R_c$ is fixed and $r_e$ is
optimized at the CAS level with tight convergence. This is followed by a FCI calculation 
to obtain the energy at this geometry, with typical errors in the CAS geometry
optimization compared to that of the FCI geometries on the order of a m\AA.  A
conical intersection between the $1^2A_1$ and $1^2B_2$ surfaces is observed at
$R_c=2.51$\AA~which is the result of Jahn-Teller splitting of the $1^2E'$ $D_{3h}$ surface \cite{gerber}.  
\begin{figure*}[!ht]
\resizebox{4in}{!}{\includegraphics{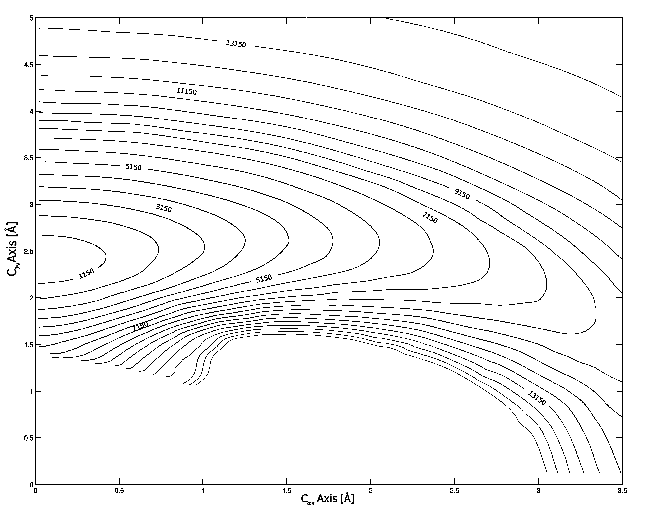}}
\caption{
\label{adp}
The Li${}_3~$ $1^2A''$ potential energy surface calculated at the 
second order spin restricted open-shell M{\o}ller-Plesset perturbation theory
for collision angles $90^\circ$ ($C_{2v}$) to $45^\circ$.  The dissociation
energy is $14156.5$cm${}^{-1}$ with $R_C=2.40$\AA~and $r_e=2.77$\AA~in
the $C_{2v}$ symmetry.}
\end{figure*}

Preliminary calculations of the Li${}_3~$ $1^2A''$ surface have been carried out using
second order spin restricted open-shell M{\o}ller-Plesset perturbation (ROMP2) theory as
implemented in Gaussian 03 \cite{gaussian}.  Here the IOM CVTZ basis set
\cite{ironmartin} was used for calculation size convenience, with a benchmark
diatomic bond length error of $0.075$\AA~at ROMP2 and $0.008$\AA~at RCCSD(T).  The
analytic potential energy surface was interpolated using cubic splines with $256$
{\em ab initio} data points and is shown in Figure \ref{adp}.  This $1^2A''$
state has a dissociation energy of
$14156.5$~cm${}^{-1}$ at the equilibrium triangular geometry of
$R_C=2.40$\AA~and $r_e=2.77$\AA~in $C_{2v}$.

\section*{Conclusions}

The electronic ground state $1^2A'$ surface of the lithium trimer has been
calculated using valence electron FCI theory with the lithium cores represented
using the SBK pseudopotential \cite{sbk}.  It was found to be necessary
to systematically include core-valence correlation in the calculation for
precise calculations of diatomic spectroscopic values. 
The basis set chosen is a $p$ and $d$ function
augmentation of the SBK basis set \cite{sbk} as given by Labello, Ferreria, and
Kuntz \cite{LFK1} with
the $s$, $p$ and $d$ functions optimized with the inclusion of a core
polarization potential to predict the correct diatomic dissociation energy.
With the recommended basis set of this work, the use of the core polarization
potential to include core-valence effects lead to an improvement $26.4$ m\AA~in
the bond length, $6.44$ cm${}^{-1}$ of the vibrational frequency and $144.605$
cm${}^{-1}$ dissociation energy of Li${}_2$.

To interpolate between sparse {\em ab initio} data points, the full interpolant
moving least squares method was implemented using a scaled exponential weighting
function with a smooth cutoff function as a constraint on the number of included
data points.  The surface was expanded using the inverse spatial coordinates
with a $6$ order bivariate polynomial.  With this interpolation method the
$1^2A'$ surface was calculated for collision angles $45^\circ$ to $90^\circ$
near the equilibrium Jacobi bond lengths of $r_e=3.218$\AA~and $R_C=2.238$\AA.
A conical intersection is found between the $1^2A_1$ and $1^2B_2$ states in
$C_{2v}$ symmetry near the equilibrium geometry of the $1^2A'$ surface.  Because
of the location of this intersection it is necessary to account for its
existence in both chemical and ultracold physics.  Also a preliminary surface
for the excited state $1^2A''$ is presented at the ROMP2 level of theory.  It is
the authors intention to continue to study the long range interactions of the
lithium trimer on the ground $1^2A'$ surface and to investigate both elastic
collisions and photoassociation of the lithium diatom-atom pair for the
formation of ultracold trimers. 

\section*{Acknowledgments}

We acknowledge Professor Frank E. Harris for his many contributions to
quantum chemistry and physics.  His computational skills and
insight into developing efficient mathematical models have greatly
benefited these fields.  One of us (HHM) would like to express our 
thanks and gratitude for many years of collaborative research with
a lifelong friend.  This research was funded in part by the U.S. Department of Energy Office of Basic Energy Sciences.

\renewcommand\bibsection{\section*{References}}

\end{document}